\documentclass[prl,preprint]{revtex4}
\usepackage{graphicx}
\usepackage{epsfig}

\newcommand{\ket}[1]{|{#1}\rangle}
\newcommand{\bra}[1]{\langle{#1}|}

\begin{document}
\noindent {\Large \sffamily \bfseries
Robust photonic entanglement distribution via state-independent encoding onto decoherence-free subspace 
}

\vspace{8mm}

\noindent
Takashi Yamamoto$^{1,2}$, Kodai Hayashi$^{1,2}$, 
\c{S}ahin Kaya \"Ozdemir$^{1,2,3}$, \\ Masato Koashi$^{1,2,3}$ \& Nobuyuki Imoto$^{1,2,3}$
\vspace{8mm}

\noindent {\itshape
$^1$
Department of Materials Engineering Science, Graduate school of Engineering Science, Osaka University, Toyonaka, Osaka 560-8531, Japan \\
$^2$CREST Research Team for Photonic Quantum Information, 4-1-8 Honmachi, Kawaguchi, Saitama 331-0012, Japan \\
$^3$SORST Research Team for Interacting Carrier Electronics, 4-1-8 Honmachi, Kawaguchi, Saitama 331-0012, Japan
}

\vspace{8mm}

\date{\today}

{\bf 
Efficient and faithful implementation of quantum information tasks, e.g., quantum computing, quantum communication and quantum metrology \cite{Bennett00N,Gisin02,Giovannetti04S}, requires robust and state-independent decoherence-suppressing measures to protect quantum information carriers. Here we present an experimental demonstration of a robust distribution scheme in which one photon of an entangled photon pair is successfully encoded into and decoded from a decoherence-free subspace (DFS) by a state-independent scheme. We achieved a high-fidelity distribution of the entangled state over fibre communication channel, and also demonstrated that the scheme is robust against fragility of the reference frame. The scheme, thanks to its state-independence, is also applicable to multipartite case where the photon to be distributed is entangled with many other photons. Such a universal scheme opens the possibility of robust distribution of quantum information among quantum communication and computing networks.
}

Encoding into a decoherence-free subspace (DFS)\cite{Lidar} is one of the effective schemes to protect quantum states against decoherence by exploiting symmetries in system-environment interactions. Let us briefly introduce the robustness of quantum states in a simple example of DFS.
Consider a quantum channel that adds a random phase shift $\varphi_{\rm S}$, transforming $\ket{\rm 0}_{\rm S}\mapsto \ket{\rm 0}_{\rm S}$ and $\ket{\rm 1}_{\rm S} \mapsto e^{i\varphi_{\rm S}}\ket{\rm 1}_{\rm S}$, where $\ket{\rm 0}_{\rm S}$ and $\ket{\rm 1}_{\rm S}$ are orthogonal states of qubit S. If the sender Alice prepares the state $\alpha\ket{\rm 0}_{\rm S}+\beta\ket{\rm 1}_{\rm S}$ and sends it to the receiver Bob through the channel, the state completely dephases to $|\alpha|^2\ket{\rm 0}_{\rm S}\bra{\rm 0}+|\beta|^2\ket{\rm 1}_{\rm S}\bra{\rm 1}$. 
Now suppose that she has another qubit $\rm S^\prime$ and sends it through the channel with phase shift $\varphi_{\rm S^\prime}$, which is correlated as  $\Delta\varphi=\varphi_{\rm S}-\varphi_{\rm S^\prime}=0$. Such phase fluctuations are referred to as {\it collective dephasing}. In this case, the states of the logical qubit spanned by \{$\ket{\tilde{0}}\equiv \ket{\rm 0}_{\rm S}\ket{\rm 1}_{\rm S^\prime}$, $\ket{\tilde{1}}\equiv \ket{\rm 1}_{\rm S}\ket{\rm 0 }_{\rm S^\prime}$\} are invariant except for a physically irrelevant global phase: the channel transforms the state $\alpha\ket{\tilde{0}}+\beta\ket{\tilde{1}}$ into $\alpha e^{i\varphi_{\rm S^\prime} }\ket{\tilde{0}}+\beta e^{i\varphi_{\rm S}}\ket{\tilde{1}}=e^{i\varphi_{\rm S}}(\alpha\ket{\tilde{0}}+\beta\ket{\tilde{1}})$.  
Thus the Hilbert subspace spanned by the logical basis $\{\ket{\tilde{0}},\ket{\tilde{1}}\}$ is a DFS for the collective dephasing. 
Since fluctuations are correlated temporally and/or spatially in many practical cases, the DFS is considered as an effective method to fight against such decoherence. The experimental efforts to demonstrate the effectiveness have also been reported in ref.\cite{Kwiat00S,Mohseni03PRL,Bourennane04PRL,Prevedel07PRL,Kielpinski01S,Viola01S,Ollerenshaw03PRL,Roos06N,Chen06PRL,Yamamoto07NJP,Banaszek04PRL}. It is worth noting that the DFS is also useful against fragility/ambiguity of the reference frame shared among the communication nodes since it effectively disturbs the state in the same way\cite{Walton03PRL, Boileau04PRLb,TYamamoto05PRL}. 
As an example, let us consider a photonic qubit, which uses the polarization states of a single photon. 
If it is sent through a polarization maintaining fibre (PMF), 
fluctuations in the birefringence of the fibre dephase the qubit on the basis \{0=H, 1=V\}, where H and V represent the horizontal and vertical polarization states. 
Since these fluctuations are relatively slow in time, two photons travelling through the PMF within a short time interval receive a collecting dephasing.
On the other hand, if this photonic qubit is transmitted to/from a satellite whose orientation around the light path is not calibrated, the lack of the reference frame effectively leads to a dephasing in the circular-polarization basis \{0=L, 1=R\}. One can then send two photons in parallel to have collective dephasing against which the DFS works well.

An implementation of the DFS scheme is relatively easy if the sender Alice knows the state to be protected. She may simply generate the corresponding state in the DFS directly. However, if a given qubit forms an entangled state together with other systems, an encoding into the DFS is required. In order to keep the entanglement intact, this encoding from the physical qubit to logical one must be state-independent. 
Now let us suppose that Alice has a photon $\rm S$ which forms a part of $n$-photon entangled state 
\begin{eqnarray}
\ket{\psi}\equiv \alpha \ket{u}\ket{\rm H}_{\rm S}-\beta\ket{v}\ket{\rm V}_{\rm S},
\label{eq:1}
\end{eqnarray} 
where $\ket{u}$ and $\ket{v}$ are states of $n-1$ photons, which may or may not be owned by Alice. She wants to send the state of photon $\rm S$ to Bob through a
collective dephasing channel.
In principle, Alice can send it faithfully if she can encode the state (\ref{eq:1}) into
$\ket{{\Psi}}\equiv \alpha \ket{u}\ket{\tilde{0}}-\beta\ket{v}\ket{\tilde{1}}$
which is immune to channel noise according to the above argument. Note that this entanglement distribution requires a {\it state-independent} encoding which acts
locally on system S while keeping the quantum correlations of
the whole system intact. However, a unitary encoding scheme which
satisfies this requirement is hard to implement using current optical technologies. 
We can, however, achieve effectively a state-independent DFS encoding and decoding in a probabilistic manner 
as follows: Alice prepares the ancillary photon $\rm S^\prime$ in $
\ket{\rm D }_{\rm S^\prime}\equiv(\ket{\rm H}_{\rm S^\prime}+\ket{\rm V }_{\rm S^\prime})/\sqrt{2}$  and sends it to Bob together with photon S.  
After receiving both photons, Bob sifts out only the states in the DFS spanned by \{$\ket{\tilde{0}}= \ket{\rm H}_{\rm S}\ket{\rm V}_{\rm S^\prime}$, $\ket{\tilde{1}}= \ket{\rm V}_{\rm S}\ket{\rm H}_{\rm S^\prime}$\}. Note that this effectively encodes the state into $\ket{{\Psi}}$, which is protected against collective dephasing. The decoding into the desired state $\ket{\psi}$ is simply performed by measuring one of the photons in $\{\ket{\rm D}, \ket{\rm \bar{D}}\equiv(\ket{\rm H}-\ket{\rm V})/\sqrt{2}\}$ basis and adding $0$ or $\pi$ phase shift depending on the measurement result. Since the initial state $\ket{\psi}$ represents arbitrary  $n$-photon states, the present scheme can be applied to transfer any polarization state of a photon without destroying the entanglement.

In experiment, we performed a distribution of 
the maximally entangled photon pair in the state 
$\ket{\phi^{-}}\equiv (\ket{\rm H}_{\rm A}\ket{\rm H}_{\rm S}-\ket{\rm V}_{\rm A}\ket{\rm V}_{\rm S})/\sqrt{2}$. The corresponding encoded state is 
$\ket{{\Phi}^{-}}\equiv (\ket{\rm H}_{\rm A}\ket{\tilde{0}}-\ket{\rm V}_{\rm A}\ket{\tilde{1}})/\sqrt{2}$.
In Fig.1, we depict the experimental scheme which uses 0.5 km 
PMF as a collective
dephasing channel for two photons sent by Alice. The optical
axis of the PMF is well adjusted so that $\ket{\rm H}$ and $\ket{\rm V}$ are faithfully transmitted, but the other states are altered because of the fluctuation of the
relative phase between $\ket{\rm H}$ and $\ket{\rm V}$. 
Alice sends photon S after photon $\rm S^\prime$ within a short time delay $\Delta t_{\rm A}\sim 3~{\rm ns}$ to ensure collective dephasing, $\Delta\varphi\sim 0$. The group-velocity difference in the
PMF adds an additional temporal delay $\tau$ between H- and
V-polarized photons. However, in our scheme
this does not cause any significant problem apart from shifting
the arrival time of the successfully postselected photon S in the
output.

On the receiving side, Bob channels photon S into short {\bf $(\cal S)$} arm and photon $\rm S^\prime$ into long {\bf $(\cal L)$} arm. Then he mixes them simultaneously by a polarizing beam splitter PBS$_{\rm B}$ for sifting out the DFS. For simplicity of the experiment, we used a nonpolarizing beam splitter BS$_{\rm B}$ for the channelling. 
The correct channelling occurs with probability 1/4, and the incorrect channelling events can be discriminated through the arrival timing at the detectors D$_{\rm X}$ and D$_{\rm Y}$. 
Therefore, in the following, we only consider the case where photons S and $\rm S^\prime$ travel arms $\cal S$ and $\cal L$, respectively. 
The length difference between the arms is adjusted almost equal to 
$\Delta t_{\rm A}$ by the mirrors (M) on a motorized stage. 
The joint state of the photons in arms $\cal S$ and $\cal L$ 
is in the Hilbert space spanned by $\{\ket{\rm H_{\rm
S}}\ket{\rm H}_{\rm S^\prime}, \ket{\rm V}_{\rm S}\ket{\rm V}_{\rm
S^\prime}, \ket{\tilde{0}}, \ket{\tilde{1}}\}$.
In order to extract the states in the DFS spanned by $\{\ket{\tilde{0}},
\ket{\tilde{1}}\}$ from this state, Bob rotates the polarization of the photon in arm $\cal L$ by 90$^\circ$ using a half wave plate (HWP), and postselects
the cases where there is one photon in each output of the
PBS$_{\rm B}$. This operation is referred to as the linear optical
quantum parity gate (QPG)\cite{Pittman01PRA}. The postselected state is local unitarily equivalent to the
encoded state $\ket{{\Phi}^{-}}$. Detecting a D-polarized
photon in mode X projects the state onto the expected entangled
state
\begin{eqnarray}
\ket{\phi^{-}}=\frac{1}{\sqrt{2}}(\ket{\rm H}_{\rm A}\ket{\rm H}_{\rm Y}-\ket{\rm
V}_{\rm A}\ket{\rm V}_{\rm Y}) \label{eq:f}
\end{eqnarray}
between Alice and Bob. Detection of a $\bar{\rm D}$-polarized
photon, on the other hand, prepares the state $\ket{\phi^{+}}$ which can be transformed into $\ket{\phi^{-}}$ by a
$\pi$-phase shifter. In the experiment, we neglected such events for
simplicity.
The successful events are postselected 
by using the time resolving coincidence detection within 2.5 ns time window at detectors D$_{\rm A}$, D$_{\rm X}$ and D$_{\rm Y}$. The HWP and the quarter wave plates (QWP) in front of the detectors D$_{\rm A}$ and D$_{\rm Y}$ are used for verification experiments.

We use spontaneous parametric down conversion (SPDC) with the photon pair generation rate $\gamma$ as the entangled photon source, and a weak coherent pulse (WCP) with an average photon number $\nu$  as the ancillary photon source. 
In order to achieve high fidelity, $\gamma$ and $\nu$ must be properly chosen. 
A three-fold coincidence detection may occur 
when Alice detects photons at D$_{\rm A}$ and Bob receives (i) one photon from SPDC and one from WCP, (ii) two photons from WCP, or (iii) two photons from SPDC. The desired event takes place if the detection is due to (i). 
The probabilities of the cases (i)-(iii) are $\mathcal{O}(\gamma \nu\eta^2)$, $\mathcal{O}(\gamma \nu^2 \eta^2)$, and $\mathcal{O}(\gamma^2 \eta^2)$, respectively, where $\eta$ is the channel transmittance. Provided that $1 \gg \nu \gg \gamma$ is satisfied, the probabilities of coincidence events due to (ii) and (iii) are negligible. In our experimental setting, we estimated $\gamma \sim 10^{-3}$ 
, and thus to satisfy the above condition we chose $\nu \sim 10^{-1}$. 

In order to demonstrate the power of our scheme, we performed experiments to see that (a) 
a highly entangled state $\rho_{\rm{AS}}$ was generated by SPDC, 
(b) 
it was degraded by the transmission channel resulting in an almost disentangled state $\rho'_{\rm{AS}}$ if our DFS method is not used, and 
(c) 
the state distributed by our DFS method was a highly entangled state $\rho_{\rm{AY}}$. 
In order to measure $\rho_{\rm{AS}}$, we blocked the paths C and $\cal L$ to stop the ancillary photon $\rm S^\prime$, and aligned the HWP in mode ${\rm X}$ such that $\ket{\rm H}$ and $\ket{\rm V}$ are not rotated.  Since all optics between D$_{\rm X}$ and the input of the PMF are considered as a polarizer, 
$\rho_{\rm{AS}}$ can be measured through basis selection by inserting a QWP and a HWP {\it before} the PMF and recording the coincidence events between D$_{\rm A}$ and D$_{\rm X}$. 
If we insert the wave plates {\it after} the PMF, we can similarly measure $\rho'_{\rm{AS}}$. We performed the measurements over 5 s in each basis for (a) and (b), and  over 800 s for (c).

Violation of Bell inequality is one of the strongest signatures of quantum correlations. 
From a set of measured polarization correlations, the violation of Clauser-Horne-Shimony-Holt (CHSH)-type Bell inequality\cite{Clauser69PRL} is tested by calculating the Bell parameter $S^{\rm CHSH}$ which is bounded by 2 in any local realistic theory. On the other hand, quantum mechanics predicts the violation of this bound up to a value of $S^{\rm CHSH}=2\sqrt{2}$ by the maximally entangled state. In the experiments, we found 
$S^{\rm CHSH}=2.36\pm0.09$ for 
$\rho_{\rm{AY}}$ (for the other states, see Table 1 ). The violation of the local realistic limit by 4 standard deviations 
implies the existence of genuine quantum correlations in 
the state distributed using our DFS scheme.

For further evaluation of our scheme, we estimated the fidelity to the state $\ket{\phi^{-}}$ and also reconstructed the density matrices of $\rho_{\rm AS}$, $\rho^{\prime}_{\rm AS}$, and $\rho_{\rm AY}$ by measuring several polarization correlations. 
The density matrices estimated using the
maximally likelihood method\cite{James01PRA} are shown in Fig.~2. The presence of
the off-diagonal elements in the density matrix is a strong sign
of the quantum coherence. It is seen that initially present
quantum coherence (Fig.2a) is significantly destroyed during
transmission in the channel (Fig.2b). On the other hand, with the application of our protocol, the quantum coherence is restored (Fig.2c). The effectiveness of the protocol is quantitatively analysed by the calculated fidelities to the desired state ($F\equiv \bra{\phi^{-}}\rho\ket{\phi^{-}}$) and the amounts of entanglement ($E$) as seen in Table 1. The slight deviation from
the ideal transmission can be attributed to the mode mismatch and
multi-photon effects, which can be minimized using advanced 
photon sources\cite{Beugnon06N,Moehring07N}. These results suggest without doubt that
our protocol provides the DFS-based entanglement protection against collective decoherence in 0.5km optical fibre.

It is important to note here that the significant advantage of our
protocol over the existing schemes\cite{{Gisin02}} is its robustness against the
fluctuations in a reference frame of distant locations. In
the conventional methods using interferometers at the sender
and receiver, sub-wavelength adjustments are required which
makes constructing quantum networks with many nodes very
difficult. However, as we have verified in experiments (Fig.3),
our scheme does not require sub-wavelength precision; it is enough
to adjust the time delay within the coherence time of the photons. In
our experiment, we measured the coherence length as 130 $\mu$m
which is about 160 times  as long as the wavelength $\lambda=$0.79
$\mu$m. This feature of our scheme is important both for quantum
communication without shared reference frame\cite{Bartlett07RMP} and for
easy-to-implement architectures for the quantum communication
networks.

In summary, we demonstrated a robust and faithful entanglement-distribution scheme which employs a state-independent encoding into DFSs. Although the demonstration was performed on a bipartite entangled state, the approach can be extended to multipartite states because of the state-independence. 
Another significant feature of our scheme is its robustness against fluctuations of reference frame among distant nodes. 
Furthermore, we should mention that the protocol can perform equally well in optical networks connected with standard single mode optical fibres\cite{TYamamoto05PRL}. 
The versatile, robust and high-fidelity entanglement distribution scheme demonstrated here has the potential to become an integral part of the future quantum computation and communication networks.

\vspace{5mm}
\noindent
{\large {\bf Acknowledgements} }

\noindent
This work was supported by 21st Century COE Program by the Japan Society for the Promotion of Science and a MEXT Grant-in-Aid for Young Scientists.

\newpage

\begin{table}[h]

\begin{ruledtabular}
\begin{tabular}{cccc}
 $$ &$S^{\rm CHSH}$ & $F$ & $E$ \\
 \hline
  $\rho_{\rm AS}$ & $2.76\pm 0.03$  &  $1.00\pm 0.03$ & $0.958\pm 0.010$  \\
  $\rho^{\prime}_{\rm AS}$ & $1.37\pm 0.04$ & $0.46\pm 0.03$    & $0.020\pm 0.017$ \\
  $\rho_{AY}$ & $2.36\pm 0.09$ &  $0.87\pm 0.07$  &  $0.60\pm 0.11$ \\
\end{tabular}
\end{ruledtabular}
\caption{{\bf Characterization of the DFS-based entanglement distribution.} 
The Bell parameter $S^{\rm CHSH}$ and the fidelity ($F$) of the states to $\ket{\phi^{-}}$ are estimated directly from the coincidence count statistics. Amount of entanglement ($E$: entanglement of formation\cite{Wootters98PRL}) of the states are obtained from the reconstructed density matrices. The error bars indicate s.d. with the assumption of the Poisson statistics of the counts, which are calculated directly for $S^{\rm CHSH}$ and $F$, and estimated through Monte Carlo simulations for $E$. 
}
\end{table}

\newpage

\begin{figure*}[htbp]
 \includegraphics[scale=1]{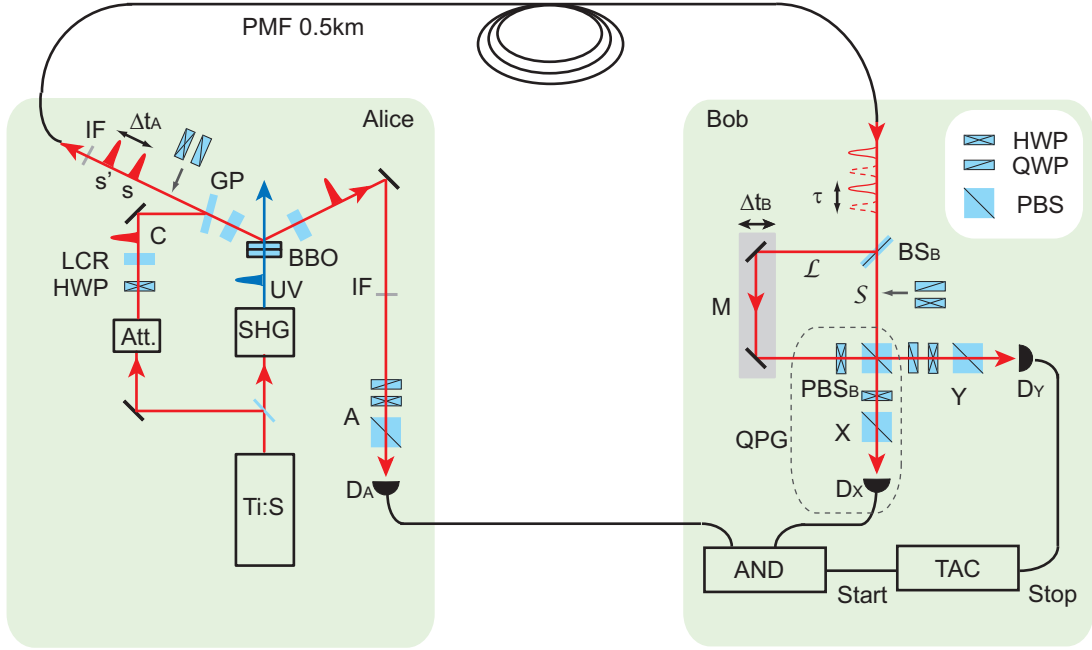}
 \caption {
\noindent {\bf Experimental set-up.} At Alice's
side, entangled photon pairs are prepared by spontaneous
parametric down-conversion (SPDC) from a pair of Type I phase
matched 1.5-mm-thick $\beta$-barium borate (BBO) crystals, which
are stacked back-to-back with their optical axes orthogonal to
each other\cite{Kwiat00S,Kim00PRA,TYamamoto03N}. Ultraviolet light pulse (average power: 50 mW,
45$^\circ$ polarization) for the pumping of the BBO crystals is prepared
by second harmonic generation (SHG) using the light pulse from
mode-locked Ti:sapphire laser (wavelength: 790 nm, pulse width: 90
fs, repetition rate: 82 MHz). The group delay between the H- and the V-polarization is compensated by inserting BBO crystals in each spatial
mode. Ancillary photons are prepared from the Ti:S laser through a combination of attenuators (Att.) and the glass plate GP (T$\sim 85 \%$ ). 
Their polarization states are set by a half wave plate (HWP) and
liquid crystal retarder (LCR). GP also serves to channel the signal (S) and the ancillary photon ($\rm S^\prime$) into the PMF. 
The spectral filtering of the generated photons is performed by a narrow band interference filter (IF, wavelength: 790 nm, band width: 2.7 nm).
At Bob's side, the received two photons are separated into short ($\cal S$) and long ($\cal L$) arms by BS$_{\rm B}$, and mixed again by the PBS$_{\rm B}$. The HWP in arm ${\cal L}$ rotates the
polarization by 90$^\circ$, while  the HWP in mode ${\rm X}$
rotates it by 45$^\circ$. The temporal delay $\Delta t_{\rm B}$ is
adjusted by the mirrors (M) on a motorized stage. The apparatus
inside the broken boxes correspond to the linear optical QPG. The verification of the shared entangled state is
performed in mode A and Y. All the detectors D$_{\rm A}$, D$_{\rm
X}$ and D$_{\rm Y}$ are silicon avalanche photodiodes placed after
single-mode optical fibres. The three-fold coincidence events are
taken by the gated discriminator (AND) and time-to-amplitude
converter (TAC). Events within 2.5 ns time window are selected as
the successful events.
}
\end{figure*}

\newpage
\begin{figure}[bhtp]
 \includegraphics[scale=1]{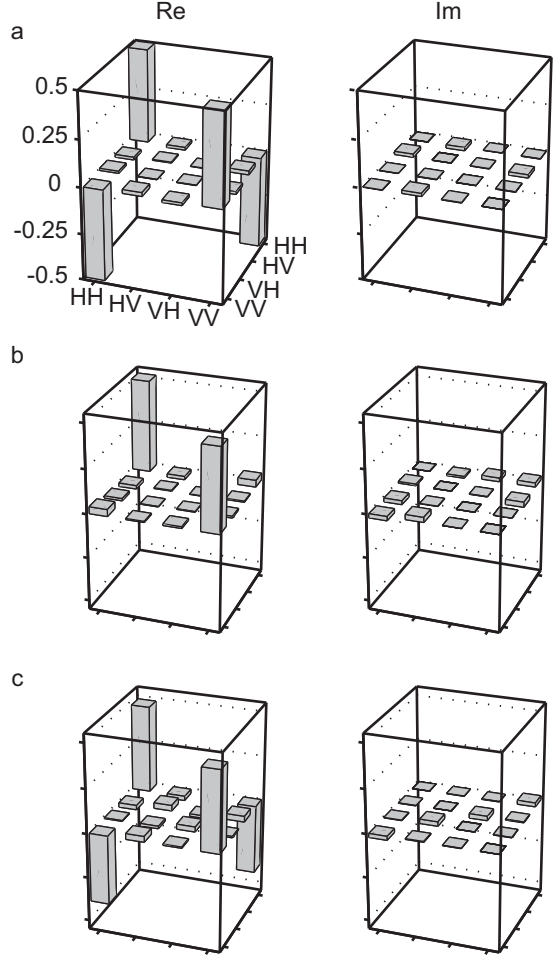}
 \caption {
\noindent  {\bf The density matrices estimated using
quantum state tomography.}
 Real and imaginary components of density matrices of {\bf a} the initially prepared state $\rho_{\rm AS}$, {\bf b} the decohered state $\rho^{\prime}_{\rm AS}$
 and {\bf c} the DFS-protected state $\rho_{\rm AY}$ are shown.  Reconstruction is done by recording coincidence counts on each of 16 different settings
 of QWP and HWP.
Total counts are 7404, 8076 and 1025, respectively for $\rho_{\rm AS}$, $\rho^{\prime}_{\rm AS}$ and $\rho_{\rm AY}$. 
}
\end{figure}

\newpage

\begin{figure}[htbp]
 \includegraphics[scale=1]{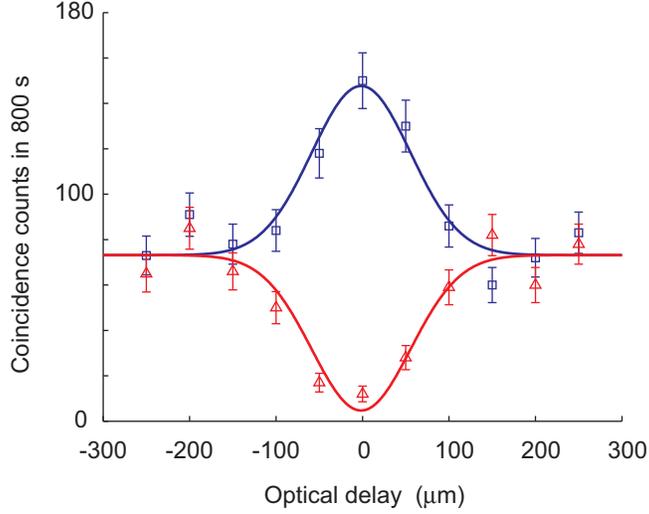}
 \caption {
\noindent 
{\bf Robustness of the scheme against path-length mismatch.}
The triangle and square markers show the coincidence counts measured on the bases $\ket{\rm D}_{\rm A}\ket{\rm D}_{\rm Y}$ (triangle) and $\ket{\rm D}_{\rm A}\ket{\bar{\rm D}}_{\rm Y}$ (square).
The optical delay is changed by moving the motorised stage M. The
calculated visibility is $0.85\pm 0.04$ at the zero delay. The
solid Gaussian curves represent the best fit to the experimental
data. The coherence length $l_c$(FWHM) is calculated as $\sim 130 \ 
\rm \mu m\simeq160 \ \lambda$. 
The error bars indicate s.d. which is calculated with the assumption of the Poisson statistics of the counts.
}
\end{figure}

\end{document}